\documentclass[aps,showpacs,twocolumn,amsmath,amssymb,superscriptaddress,prl]{revtex4}

\usepackage{graphicx}
\usepackage{bm}
\usepackage[dvips]{color} 

\begin{document}

\title{Spin Pumping Driven by Bistable Exchange Spin Waves}

\author{K. Ando}
\email{ando@imr.tohoku.ac.jp}
\affiliation{Institute for Materials Research, Tohoku University, Sendai 980-8577, Japan}
     
\author{E. Saitoh}
\affiliation{Institute for Materials Research, Tohoku University, Sendai 980-8577, Japan}
\affiliation{CREST, Japan Science and Technology Agency, Sanbancho, Tokyo 102-0075, Japan}
\affiliation{The Advanced Science Research Center, Japan Atomic Energy Agency, Tokai 319-1195, Japan }

\date{\today}

\begin{abstract}
Spin pumping driven by bistable exchange spin waves is demonstrated in a Pt/Y$_3$Fe$_5$O$_{12}$ film under parametric excitation. In the Pt/Y$_3$Fe$_5$O$_{12}$ film, the spin pumping driven by parametric excitation selectively enhances the relaxation of short-wavelength exchange spin waves, indicating strong coupling between the exchange spin waves and spin currents at the interface through efficient spin transfer. The parametric spin pumping, furthermore, allows direct access to nonlinear spin wave dynamics in combination with the inverse spin Hall effect, revealing unconventional bistability of the exchange spin waves. 
\end{abstract}


\maketitle

Nonlinear phenomena have been essential for exploring physics and technology in condensed matter. Nonlinear dynamics is ubiquitous in nature, including mechanical~\cite{Landau}, optical~\cite{Gibbs}, and magnetic~\cite{Mayergoyz} systems. This plays a significant role also in the field of spintronics, where magnetization dynamics is coupled with spin currents. The coupling between dynamical magnetization and spin currents gives rise to generation of spin currents from precessing magnetization in nonmagnetic/ferromagnetic junctions: the spin pumping~\cite{Tserkovnyak1,Brataas,Mizukami,Heinrich,Costache,PhysRevB.82.214403,kajiwara,PhysRevLett.107.046601,PhysRevB.83.144402,PhysRevLett.107.066604}. In particular, the spin pumping in metal/magnetic insulator junctions offers a route for exploring nonlinear spin physics owing to the exceptionally low magnetic damping.

Due to the nonlinearlity of magnetization dynamics, parametric excitation of dynamic magnetization is possible without fulfillment of ferromagnetic resonance (FMR) conditions, provided that the amplitude of a pumping microwave magnetic field $h$ is large enough to overcome the spin wave relaxation. Recent studies have revealed that the parametrically-excited short-wavelength spin waves drive the spin pumping in metal/magnetic insulator junctions~\cite{PhysRevLett.106.216601,ando:092510,kurebayashi:162502}; spin waves parametrically excited by a microwave in the magnetic insulator induce a spin current in the adjacent metal, giving rise to an electric voltage through the inverse spin Hall effect (ISHE) in the metal. Thus the spin pumping allows not only to generate spin currents in a wide range of materials~\cite{AndoNMat,kajiwara} but also to explore interaction between spin currents and dipolar/exchange spin waves using the ISHE~\cite{PhysRevLett.106.216601,KurebayashiNMat,ando:092510,kurebayashi:162502}.

In this Letter, we demonstrate spin pumping driven by bistable exchange spin waves. We generate two types of parametric spin waves, small-wavevector dipolar-exchange spin waves and large-wavevector exchange spin waves, through three-magnon splitting by controlling the strength of an external static magnetic field. The exchange spin waves are found to couple strongly with spin currents through efficient spin transfer due to spin pumping in a metal/magnetic insulator junction. Furthermore, using the spin pumping driven by the exchange spin waves in combination with the ISHE, we found that only exchange spin waves show hysteresis with respect to an external magnetic field and microwave excitation power, demonstrating bistable excitation of spin waves due to a negative nonlinear damping.

The sample used in this study is a Pt/La-substituted Y$_{3}$Fe$_{5}$O$_{12}$ (Pt/La:YIG) bilayer film. The single-crystal La:YIG (111) film [$2\times 2$ mm$^{2}$] with a thickness of 2 $\mu$m was grown on a Gd$_{3}$Ga$_{5}$O$_{12}$ (111) substrate by liquid phase epitaxy, where La was substituted to match the lattice constant between the film and the substrate. The 10-nm-thick Pt layer was then sputtered in an Ar atmosphere on the top of the film and two electrodes are attached to the edges of the Pt layer to detect a spin current generated by the spin pumping; a spin current generated by the spin pumping is converted into an electric voltage $V$ through the ISHE in the Pt layer~\cite{Saitoh,KimuraPRL,Valenzuela,Seki,AndoPRB}. Here, for the measurement, the Pt/La:YIG film was placed at the center of a TE$_{011}$ cavity, where a microwave with the frequency $f=9.44$ GHz exists. The background voltage $V_\text{b}$ was subtracted from the measured voltage $\bar{V}$: $V=\bar{V}-V_\text{b}$. An external magnetic field ${\bf H}$ was applied along the film plane perpendicular to the direction across the electrodes. All measurements were performed at room temperature.

\begin{figure}[bt]
\includegraphics[scale=1]{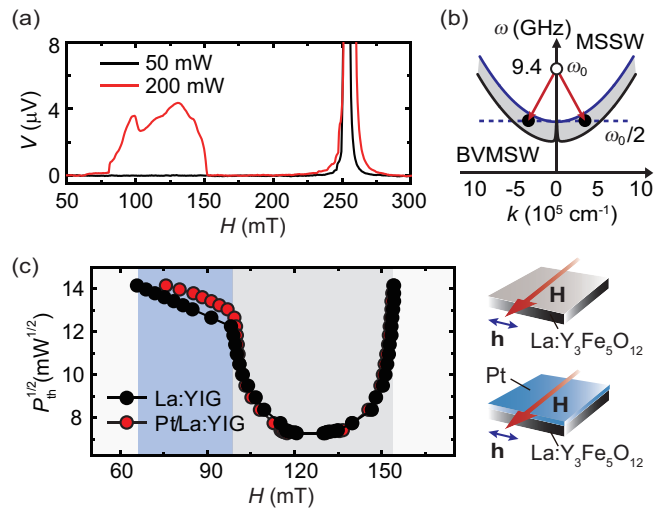}
\caption{(a) Field $H$ dependence of an electric voltage $V$ at microwave excitation powers $P=50$ and 200 mW for the Pt/La:YIG film. (b) The spin wave dispersion for the Pt/La:YIG film when the parametric excitation condition. (c) The square root of the threshold power $P_\text{th}$ for the La:YIG and Pt/La:YIG films obtained from the microwave absorption signal. }
\label{fig1} 
\end{figure}

In Fig.~\ref{fig1}(a), we show the $V$ signals measured for the Pt/La:YIG film at $P=50$ and 200 mW microwave excitation powers. Figure~\ref{fig1}(a) shows that a voltage signal appears around the ferromagnetic resonance (FMR) field $H_\text{FMR}\sim 260$ mT. This is the electric voltage induced by the ISHE and spin pumping driven by the uniform magnetization precession~\cite{ando:092510}. Notable is that an additional $V$ signal appears far below the FMR field [see $H\sim 130$ mT] only for $P=200$ mW. This additional voltage signal is induced by the spin pumping driven by parametric spin wave excitation~\cite{ando:092510,kurebayashi:162502}. Spin waves with the wavevector of ${\bf k}_p \neq 0$ can be excited through a three-magnon splitting process; when the microwave magnetic field with frequency $\omega_0=2\pi f$ is applied perpendicular to the static magnetic field ${\bf H}$, a pair of modes with the wavevectors of ${\bf k}_p$ and ${\bf -k}_p$ are excited parametrically via a ${\bf k=0}$ virtual state when a pair of modes are fed sufficient energy to overcome the dissipation, i.e., when the excitation microwave power $P$ exceed the threshold power $P_\text{th}$ [see Fig.~\ref{fig1}(b)]~\cite{Turbulence,Suhl,Wigen,PhysRevLett.72.2093}. The energies of the parametrically excited spin waves, $\omega_{{\bf k}_p}$ and $\omega_{-{\bf k}_p}$, are $\omega_0/2$ because of the energy conservation $\omega_0=\omega_{{\bf k}_p}+\omega_{-{\bf k}_p}$ and $\omega_{{\bf k}_p}=\omega_{-{\bf k}_p}$.

The spin pumping from parametrically excited spin waves strongly affects their damping. Figure~\ref{fig1}(c) shows the $H$ dependence of $P_\text{th}^{1/2}$ for the Pt/La:YIG film and a La:YIG film, where the Pt layer is missing. Here, the spin-wave damping is proportional to the square root of the threshold power $P_\text{th}^{1/2}$ of parametric excitation~\cite{Wigen}. $P_\text{th}^{1/2}$ was obtained not by the voltage measurements but by monitoring the microwave absorption signal to compare $P_\text{th}$ for the Pt/La:YIG film and that for the La:YIG film; we measured the microwave absorption signal using a field lock-in technique~\cite{ando:092510} and defined $P_\text{th}$ as the minimum microwave power at which the absorption signal becomes nonzero~\cite{lockin}. In Fig.~\ref{fig1}(c), no difference is found in the threshold values for the Pt/La:YIG and La:YIG films in the gray area. This shows that the applied microwave power is identical for these films. In contrast, notably, the threshold values are increased by attaching the Pt layer in the blue area, indicating that the spin-wave damping is enhanced in this field range. Notable is that, in the gray and blue areas, different types of spin waves are excited. These spin waves have been assigned to small-wavevector $k_p\sim 5\times 10^4$ cm$^{-1}$ dipolar-exchange spin waves with $\theta_{k_p}\sim 0-90^\circ$ [the gray area] and large-wavevector $k_p\sim 3\times 10^5$ cm$^{-1}$ exchange spin waves with $\theta_{k_p}\sim 45^\circ$ [the blue area], where $\theta_{k_p}$ is the angle between the external magnetic field ${\bf H}$ and the wavevector ${\bf k}_p$ of the spin waves~\cite{patton,PhysRevB.51.15085}.

The above experimental results suggest the strong coupling between the short-wavelength exchange spin waves and spin currents through the spin pumping. The spin pumping is known to enhance the relaxation of uniform magnetization precession, or the ${\bf k}={\bf 0}$ mode, because the spin-current emission from a ferromagnetic film deprives the magnetization of the spin-angular momentum, giving rise to an additional damping~\cite{Tserkovnyak1,Mizukami}. In the Pt/La:YIG film, however, we found that the change of the Gilbert damping $\alpha$ due to the spin pumping is negligibly small by comparing the FMR spectra for the La:YIG and Pt/La:YIG films. This is attributed to the thick, 2 $\mu$m, La:YIG layer, since the relaxation enhancement is inversely proportional to the thickness of the ferromagnetic layer~\cite{Mizukami}. The experimental result shown in Fig.~\ref{fig1}(c) indicates that this is also the case for the dipolar-exchange spin waves; the relaxation of the dipolar-exchange spin waves is insensitive to the spin-current emission. In contrast, the exchange spin waves with $\theta_{k_p}\sim45^\circ$ excited through the three-magnon interaction may tend to localize near the interface in the length scale of $1/k_p\sim 100$ nm because of the large-wavevector $k_p$. This induces efficient transfer of spin-angular momentum from the exchange spin waves to spin currents at the Pt/La:YIG interface, giving rise to significant enhancement of the relaxation.

\begin{figure}[bt]
\includegraphics[scale=1]{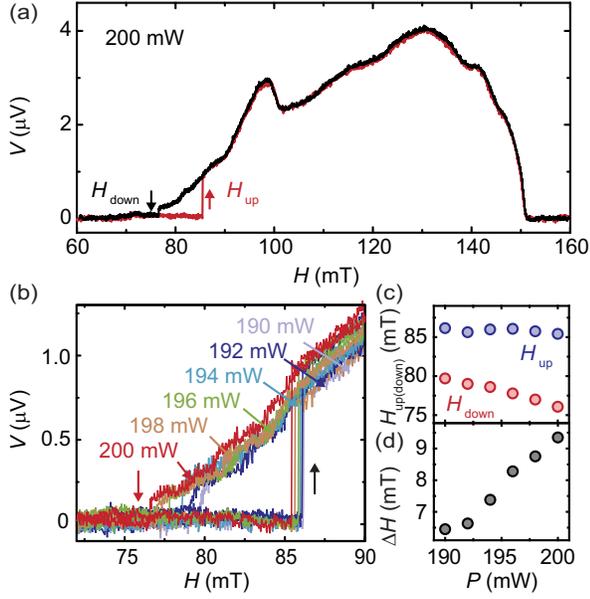}
\caption{(a) $H$ dependence of $V$ for the Pt/La:YIG film at $P=200$ mW. The red curve was measured with increasing $H$ (up sweep). The black curve was measured with decreasing $H$ (down sweep). (b) The hysteresis of $V$ with respect to $H$ measured at different microwave excitation powers for the Pt/La:YIG film. (c) $P$ dependence of $H_\text{up}$ and $H_\text{down}$, where $H_\text{up}$ and $H_\text{down}$ represent the ``jump" field of $V$ for the up sweep and down sweep, respectively. (d) $P$ dependence of $\Delta H=H_\text{up}-H_\text{down}$. }
\label{fig2} 
\end{figure}

\begin{figure}[tb]
\includegraphics[scale=1]{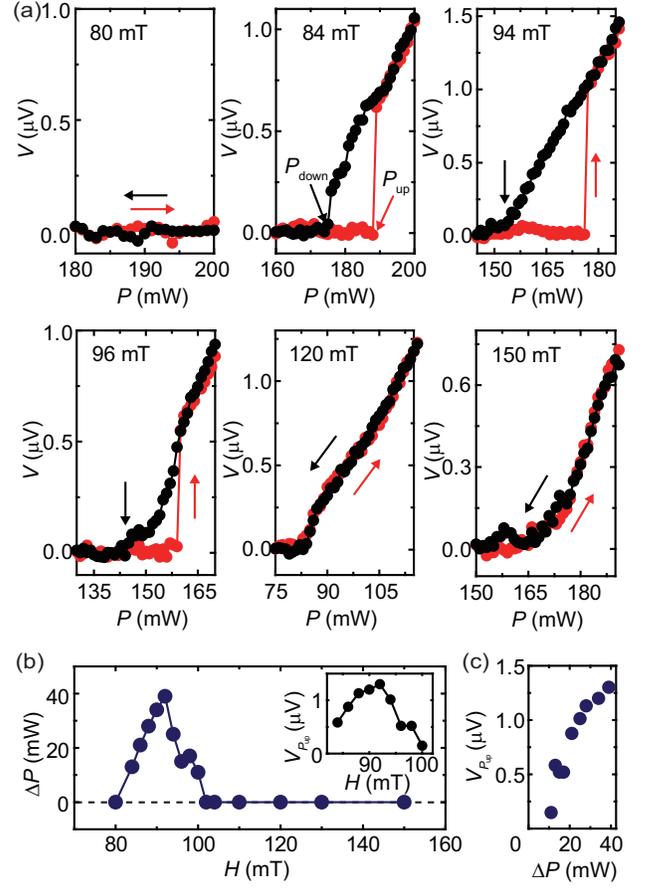}
\caption{
(a) $V$ as a function of $P$ for the Pt/La:YIG film at $H=80-150$ mT. The red(black) circles were measured with increasing(decreasing) $P$. (b) $H$ dependence of the width of the hesterysis $\Delta P=P_\text{up}-P_\text{down}$, where $P_\text{up(down)}$ denotes the microwave power where $V$ jumps as shown in the $V$ data at $H=84$ mT. The inset shows $H$ dependence of $V_{P_\text{up}}$, where $V_{P_\text{up}}$ represents the magnitude of the electric voltage at $P_\text{up}$. (c) $\Delta P$ dependence of $V_{P_\text{up}}$. 
} 
\label{fig3} 
\end{figure}

The exchange spin waves further show a nontrivial feature: bistability. Figure~\ref{fig2}(a) shows the $V$ signals measured at $P=200$ mW by sweeping the external magnetic field from 60 to 160 mT [the red curve] or from 160 to 60 mT [the black curve]. In Fig.~\ref{fig2}(a), a ``jump" in the $V$ spectrum is observed around $H_\text{up}=85$ mT when $H$ is increased from 60 mT. By decreasing $H$, the ISHE voltage shows hysteresis; the ``jump" field, $H_\text{up}$ and $H_\text{down}$, depends on the history of the field sweep as indicated by the black and red arrows, demonstrating bistability of the parametrically excite spin waves. The $V$ signals at different microwave excitation powers are shown in Fig.~\ref{fig2}(b). Figure~\ref{fig2}(b) shows systematic variation of the jump fields $H_\text{up}$ and $H_\text{down}$ [see also Fig.~\ref{fig2}(c)]. The difference between $H_\text{up}$ and $H_\text{down}$, $\Delta H=H_\text{up}-H_\text{down}$ increases with $P$ [see Fig.~\ref{fig2}(d)], a feature similar to the foldover measured at FMR~\cite{PhysRevB.80.184422}. However, the origin of the bistability of the parametrically excited spin waves is different from that of FMR. The bistability of FMR is induced by the reduction of a demagnetization field due to high power excitation; the bistability appears in the field range where the number of excited magnons is large enough to change the resonance condition~\cite{Gurevich}. Thus bistability of parametrically excited spin waves is also expected to appear for fields where $V$ is large, i.e., for dipolar-exchange spin waves, which is in contrast to the observation: bistability of the exchange spin waves. Here, the Pt layer is not essential for the bistability; bistable states appear also in the La:YIG film, which was confirmed by monitoring the microwave absorption signal.

We further show the evidence that bistable states appear only for the exchange spin waves. Figure~\ref{fig3}(a) shows the $V$ signals measured with increasing $P$ (the red circles) or decreasing $P$ (the black circles) for different external magnetic fields $H$. Clear bistable states are observed at $H=84$, 94, and 96 mT. In contrast, at higher magnetic fields, 120 mT and 150 mT, we found no hysteresis in the $V$ signal. As shown in Fig.~\ref{fig3}(a), the appearance of the bistable state is irrelevant to the threshold powers; the external magnetic field strength is essential for the bistability. The width of the hysteresis $\Delta P=P_\text{up}-P_\text{down}$ [see Fig.~\ref{fig3}(a) for 84 mT] for different external magnetic fields $H$ is plotted in Fig.~\ref{fig3}(b). Figure~\ref{fig3}(b) shows that the bistable state appears only in the magnetic field range of $80-100$ mT, where large-wavevector exchange spin waves are excited [see Fig.~\ref{fig1}(c)].

The observed bistability can be attributed to a negative nonlinear damping of parametrically excited spin waves; the damping of these spin waves decreases with increasing the number $n_{{\bf k}_{p}}$, $\partial \eta_{{\bf k}_{p}}/\partial n_{{\bf k}_{p}}<0$, where $\eta_{{\bf k}_{p}}$ is the damping of the parametrically excited spin waves. The damping of a spin wave with a negative nonlinear damping can be expressed as $\eta_{{\bf k}_{p}}=\eta_0-\eta_1 n_{{\bf k}_{p}}$. The negative nonlinear damping is negligible below $P_\text{up}$ when $P$ is increased from 0, since $n_{{\bf k}_{p}}$ is negligibly small in this situation. Thus by increasing $P$, parametric spin waves are excited above the threshold for a system without the negative nonlinear damping~\cite{Gurevich} $P_\text{up}=\eta_{{\bf k}_{p}}/G_{{\bf k}_{p}}=\eta_0/G_{{\bf k}_{p}}$, where $G_{{\bf k}_{p}}$ is the coupling parameter. In a system without the negative nonlinear damping, $n_{{\bf k}_{p}}=0$ at $P_\text{up}$~\cite{Turbulence}. The situation changes drastically by taking into account the negative nonlinear damping; assuming the relation between $n_{{\bf k}_{p}}$ and $P$ above threshold power $P_\text{th}$ as $n_{{\bf k}_{p}}=n_0\sqrt{P-P_\text{th}}$, we find $n_{{\bf k}_{p}}=0$ and $2 n_0^2 \eta_0 \eta_1/(V^2+n_0^2 \eta_1^2)$ at $P_\text{up}$; a bistable state of the parametric spin waves appears at $P_\text{up}$. Since the existence of the parametric spin waves reduces their damping when $\eta_1\neq 0$, the threshold power when $P$ is decreased from $P_\text{up}$, $P_\text{down}=\eta_{{\bf k}_{p}}/V$, is smaller than $P_\text{up}$ and thus hysteresis appears in the $V$ signal.

The negative nonlinear damping arises from a three magnon confluence process, where a parametrically excited spin wave with the wavevector ${\bf k}_{p}$ is annihilated together with a thermal spin wave with ${\bf k}_1$, creating another thermal spin wave with ${\bf k}_2$: ${{\bf k}_{p}}+{{\bf k}_1}={{\bf k}_2}$ and $\omega_{{\bf k}_{p}}+\omega_{{\bf k}_1}=\omega_{{\bf k}_2}$. The damping of the parametrically excited spin waves due to this three magnon confluence process is expressed as~\cite{Zakharov} 
\begin{equation}
\eta^c_{{\bf k}_{p}}\propto 4\pi\int (n_{{\bf k}_1}-n_{{\bf k}_2})\delta({{\bf k}_{p}}+{{\bf k}_1}-{{\bf k}_2})\delta(\omega_{{\bf k}_{p}}+\omega_{{\bf k}_1}-\omega_{{\bf k}_2})d{\bf k}_1 d{\bf k}_2, \label{damping}
\end{equation}
where $n_{{\bf k}_1}$ and $n_{{\bf k}_2}$ are the number of thermal spin waves with the wavevector of ${\bf k}_1$ and ${\bf k}_2$. When $n_{{\bf k}_1}$ and $n_{{\bf k}_2}$ are close to their thermodynamic equilibrium values, $\bar{n}_{{\bf k}_1}$ and $\bar{n}_{{\bf k}_2}$, the damping due to the three magnon confluence process can be expressed as $\eta^c_{{\bf k}_{p}}=c(\bar{n}_{{\bf k}_1}-\bar{n}_{{\bf k}_2})\equiv \eta_{{\bf k}_p}^l$, where $c$ is a proportionality constant; the difference in the thermodynamic equilibrium number of the thermal spin waves leads to a positive linear damping. However, the three magnon confluence process increases $n_{{\bf k}_2}$ and reduces $n_{{\bf k}_1}$. Thus this process reduces $n_{{\bf k}_1}-n_{{\bf k}_2}$, whereas relaxation processes of the thermal spin waves tend to return this difference to the thermodynamic equilibrium value. This competition is expressed as rate equations, $\dot{n_{{\bf k}_1}}=-\eta_{{\bf k}_1} (n_{{\bf k}_1}-\bar{n}_{{\bf k}_1})-\eta^c_{{\bf k}_p}n_{{\bf k}_p}$ and $\dot{n_{{\bf k}_2}}=-\eta_{{\bf k}_2} (n_{{\bf k}_2}-\bar{n}_{{\bf k}_2})+\eta^c_{{\bf k}_p} n_{{\bf k}_p}$. In the equilibrium condition, $\dot{n_{{\bf k}_1}}=\dot{n_{{\bf k}_2}}=0$, when $c n_{{\bf k}_p}(\eta_{{\bf k}_1}^{-1}+\eta_{{\bf k}_2}^{-1})\ll1$, the damping of the parametrically excited spin waves due to the confluence process is given by $\eta^c_{{\bf k}_p}=\eta_{{\bf k}_p}^l-\eta_{{\bf k}_p}^n n_{{\bf k}_p}$, where $\eta_{{\bf k}_p}^n =
[\eta_{{\bf k}_p}^lc(\eta_{{\bf k}_1}^{-1}+\eta_{{\bf k}_2}^{-1})]$. The second term represents the negative nonlinear damping. For the confluence process of parametrically excited spin waves with small $k_p$, ${ k}_1$ and ${ k}_2$ must be very large due to the energy and momentum conservation laws. However, the damping of spin waves, $\eta_{{\bf k}_1}$ and $\eta_{{\bf k}_2}$, increases with increasing the wavevector; the negative nonlinear damping is inefficient for small $k_p$ spin waves, which is consistent with the experimental observation; bistability appears only for the large-wavevector exchange spin waves.

Figure~\ref{fig3}(c) shows the relation between the magnitude of the electric voltage at $P_\text{up}$, $V_{P_\text{up}}$, and the width of the hysteresis $\Delta P$. This result indicates that the decrease of $\Delta P$ with decreasing $H$ below 90 mT shown in Fig.~\ref{fig3}(b) is due to the reduction of $V_{P_\text{up}}$, or the number of parametrically excited spin waves at $P_\text{up}$. The origin of this reduction is the suppression of the negative nonlinear damping. The negative nonlinear damping can be suppressed for large $k_p$ spin waves due to three magnon splitting process, where a parametrically excited spin wave with ${\bf k}_p$ splits into two thermal spin waves with ${\bf k}_1$ and ${\bf k}_2$: ${\bf k}_p={\bf k}_1+{\bf k}_2$ and $\omega_{{\bf k}_{p}}=\omega_{{\bf k}_1}+\omega_{{\bf k}_2}$, since the splitting process leads to a positive nonlinear damping~\cite{Zakharov}. This process is important only for sufficiently large-wavevector spin waves; the splitting process is allowed for $k_p>k_{m}$ because of the energy and momentum conservation. Here $k_m$ is of the order of $10^5$ cm$^{-1}$~\cite{Zakharov}, indicating that the splitting process is essential for exchange spin waves. In the present system, the minimum wavevector $k_m$ is obtained from the relation $\omega_{{\bf k}_m} =2\omega_{{\bf k}_m/2}$ for the approximated dispersion relation~\cite{Sparks} $\omega_k=\gamma H +D k^2 +(1/2)\gamma (4\pi M_s)\sin^2 \theta_{k_p}$ with $\theta_{k_p} =\pi/4$ as $k_m=\sqrt{\gamma(4H+4\pi M_s)/2D}$. This indicates that the minimum wavevector $k_m$ decreases with decreasing the external field $H$, whereas the wavevector of parametrically excited spin waves $k_p$ increases with decreasing $H$ [see Fig.~\ref{fig1}(b)]; the wavevector of parametrically excited spin waves $k_p$ exceeds $k_m$ below 90 mT, which suppresses the negative nonlinear damping through the three magnon splitting process.

In summary, we demonstrated strong coupling between spin currents and short-wavelength exchange spin waves at a Pt/Y$_3$Fe$_5$O$_{12}$ interface using parametric excitation. This strong coupling is responsible for the spin pumping driven by parametrically excited spin waves, which, in combination with the inverse spin Hall effect, allows direct access to bistable states of the exchange spin waves. Further studies, such as time evolution of the bistable states for different fields and microwave powers, will provide crucial piece of information for understanding the bistability. Thus the combination of the inverse spin Hall effect and spin pumping driven by parametric spin waves promises a significant progress for understanding the physics of nonlinear spin-wave dynamics and is of crucial importance for further development of spin-wave-based spintronic devices.

The authors thank to T. An and H. Kurebayashi for valuable discussions. This work was supported by the Cabinet Office, Government of Japan through its ``Funding Program for Next Generation World-Leading Researchers," the Asahi Glass Foundation, the Sumitomo Foundation, Research Foundation for Materials Science, and JST-CREST ``Creation of Nanosystems with Novel Functions through Process Integration".


\begin{thebibliography}{10}
\bibitem{lockin}
The modulation field for the lock-in detection was 0.01 mT, which is small
  enough so that experimental artifacts due to hysteresis of exchange spin
  waves for determining the threshold are negligible. We plotted the minimum
  and maximum fields within which the absorption signal is nonzero as a
  function of the applied microwave power, which provides the relation between
  $P_\text{th}$ and $H$.
\end{thebibliography}
\end{document}